\documentclass[aps,prl,twocolumn,floatfix,showpacs]{revtex4}
\usepackage[dvips]{graphicx}
\usepackage{array}
\usepackage{bm,amssymb,amsmath}

\begin{document}


\title{Entropic analysis of quantum phase transitions  
from uniform to spatially inhomogeneous phases}

\author{\"O.~Legeza, J.~S{\'o}lyom}
\affiliation{Research Institute for Solid State Physics and Optics, H-1525
Budapest, P.\ O.\ Box 49, Hungary}
\author{L. Tincani, R.~M.~Noack}
\affiliation{Fachbereich Physik, Philipps-Universit\"at Marburg,
D-35032 Marburg, Germany}

\date{\today}

\begin{abstract}
We propose a new approach to study quantum phase transitions in
low-dimensional fermionic or spin models that go
from uniform to spatially inhomogeneous phases such as dimerized,
trimerized, or incommensurate phases. It is based on studying the 
length dependence of the von Neumann 
entropy and its corresponding Fourier spectrum for finite segments in
the ground state of finite chains.
Peaks at a nonzero wave vector
are indicators of oscillatory
behavior in decaying correlation 
functions.
\end{abstract}

\pacs{71.10.Fd, 71.30.+h, 75.10.Jm}

\maketitle

Recently, it has been shown \cite{wu} that, quite generally,
discontinuities in some measure of entanglement between different
parts of a ground-state quantum system  
can be used to study quantum phase transitions (QPTs).  
This measure can be the
concurrence \cite{wootters}, which can be used to study spin-1/2 
models \cite{osborne,osterloh,syljuasen,gu,vidal,roscilde,yang}, the entropy of
a single site \cite{zanardi,gu2,larsson_single}, the entropy of a pair of neighboring sites \cite{legeza_qpt},
or, more generally, the entropy of a block of length $l$ \cite{legeza_qpt,deng}.

In a parallel development, it has been pointed out that critical and noncritical
systems behave differently \cite{vidal_sl,korepin_sl} when the
length dependence of the entropy of a finite segment of a quantum system is studied. The von Neumann entropy of a 
subsystem of length $l$ (measured here and subsequently in units of
the lattice constant $a$)
saturates at a finite value when the system is noncritical, i.e., when
the spectrum is gapped, 
while it increases logarithmically for critical, gapless systems. An analytic expression has 
been derived for models that
map to a conformal field theory \cite{holzhey}, and
this form has been shown to be satisfied by critical spin models. 
The entropy for a subsystem of length $l$ in a finite system of length $N$
has been shown to be \cite{cardy}
\begin{equation}
  s(l) = \frac{c}{6}\ln \left( \frac{2N}{\pi} \sin \left( \frac{\pi l}{N} \right) \right)
 + g \,,
\label{eq:cardy}
\end{equation}
where $c$ is the central charge and $g$ is a constant shift due to the
open boundary which depends on the ground-state degeneracy \cite{affleck}.

The aim of this paper is to show that the length dependence of the von Neumann 
entropy of a subsystem can, in fact, display a much richer structure than discussed until now, 
and that its analysis allows a better characterization of the QPT, thereby providing a new 
diagnostic tool to study QPTs. 
Moreover, this method seems to be appropriate to 
study cases when no true phase transition takes place, i.e., when only
the character of the decaying correlation function changes.
Our method is especially convenient when the density-matrix 
renormalization-group (DMRG) algorithm \cite{white} is used, in which
the density 
matrix of blocks of different lengths are generated in the course of
the procedure so that the von Neumann
entropy can be easily calculated.
 
\begin{figure}[htb]
\includegraphics[scale=0.5]{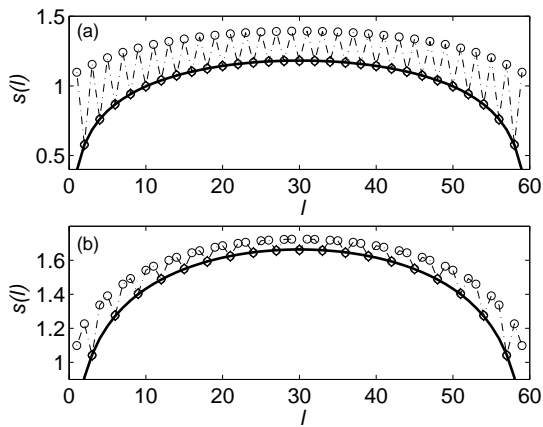}
\caption{Length dependence of the von Neumann entropy of segments of length $l$
of a finite chain with $N=60$ sites for (a) the Takhtajan-Babujian and
(b) the Uimin-Lai-Sutherland
models. The solid lines are our fit using Eq.~(\ref{eq:cardy}) taking every second and third data points.} 
\label{fig:TB-ULS}
\end{figure}

The first, simplest
case which we will consider is the Majumdar-Ghosh (MG) model
\cite{majumdar}, which corresponds to the  
frustrated spin-$1/2$ Heisenberg chain, described by the Hamiltonian 
\begin{equation}  
        {\cal H} = \sum_i \big[ J ( \bm{S}_i \cdot \bm{S}_{i+1}) +
                          J'( \bm{S}_i \cdot \bm{S}_{i+2}) \big] \, ,
\label{eq:ham_frust}
\end{equation}  
at the particular parameter value $J'=0.5 J$.
Since the ground state (of a finite chain)
consists of independent nearest-neighbor singlets 
between odd and even sites, the entropy of a segment of length $l$
oscillates between $\ln 2$ when the block has an unpaired spin at its end and 
$0$ when all spins in the block are paired into singlets \cite{majumdar}.
We will also consider the exactly solvable Takhtajan-Babujian (TB)
\cite{TB} and  
Uimin-Lai-Sutherland (ULS) \cite{ULS} points of the spin-one
bilinear-biquadratic model \cite{biqu}, 
\begin{equation}  
        {\cal H} =  \sum_i \big[ \cos\theta ( \bm{S}_i \cdot \bm{S}_{i+1}) +
        \sin\theta ( \bm{S}_i \cdot \bm{S}_{i+1})^2 \big] \,,
\label{eq:ham_biqu}
\end{equation}
which correspond to the parameter values $\theta = -\pi/4$ and
$\pi/4$, respectively. 
As shown in Fig.~\ref{fig:TB-ULS}, 
a periodic oscillation is superimposed onto a curve that is
described by the analytic form given by Eq.\ (\ref{eq:cardy}) in both
cases. 
At the TB point,
the period of oscillation is two lattice sites, while at the ULS point 
it is three lattice sites.
When the length $l$ is taken to be a multiple of two for the TB point
or a multiple of three for the ULS, the entropy
$s(l)$ can be well-fitted
using Eq.~(\ref{eq:cardy}) with $c$ approaching the known values, 
$c=3/2$ \cite{c_tb} and $c=2$ \cite{c_ls}, respectively, in the limit
of large $N$. 

\begin{figure}[htb]
\includegraphics[scale=0.5]{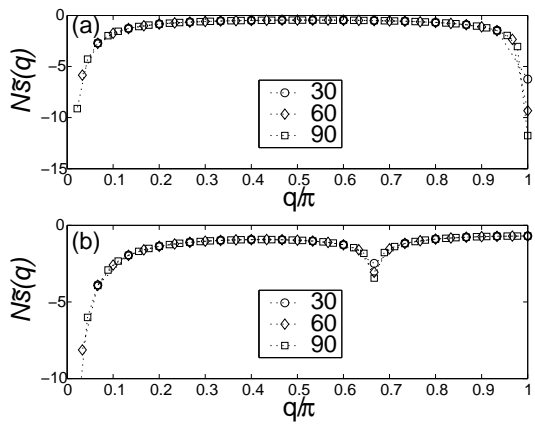}
\caption{Fourier spectrum $\tilde{s}(q)$ (scaled by the system
  size $N$) of the length-dependent von Neumann entropy of finite 
chains of length $N=30,60$ and 90 for (a) the Takhtajan-Babujian and
(b) the Uimin-Lai-Sutherland models.} 
\label{fig:TB-ULS-q}
\end{figure}

In order to analyze the oscillatory nature of the finite subsystem
entropy $s(l)$, it is
useful to consider its Fourier spectrum  
\begin{equation}
      \tilde{s}(q) = \frac{1}{N}\sum_{l=0}^N e^{- i q l}s(l)\,,
\end{equation} 
with $s(0)=s(N)=0$ where $q=2\pi n/N$ and $n=-N/2,\dots,N/2$. 
Since $s(l)$ satisfies the
relation $s(l)=s(N-l)$, its Fourier components are all real and symmetric,
$\tilde{s}(q)=\tilde{s}(-q)$; therefore, only the 
$0\le q\leq \pi$ region will be shown.
Except for the large positive $\tilde{s}(q=0)$ component that grows with increasing
chain length, the other components are all negative. They are shown for the 
two cases discussed above in Fig.~\ref{fig:TB-ULS-q}.
As can be seen, apart from the $q=0$ point, the Fourier spectrum
exhibits (negative) peaks at  
$q=\pi$ and $q=2\pi/3$, respectively.  
This is related to the fact that the TB model has two soft modes, at
$q=0$ and $\pi$, while the ULS model has three, at $q=0$ and $\pm 2\pi/3$.
Although finite-size extrapolation shows that these components vanish in the
$N \rightarrow \infty$ limit, these peaks in the Fourier spectrum
are nevertheless indications that the decay of correlation functions
is not simply algebraic  in these critical models, but that the
decaying function is multiplied by an oscillatory factor.
When the same calculation is performed for $\theta$ in the range
$-3\pi/4<\theta<-\pi/4$,
where the system is gapped and dimerized, the
peak at $q=\pi$
remains finite as $N \rightarrow \infty$. 
On the other hand, in the whole interval
$\pi/4\le\theta<\pi/2$, where the system is gapless and the excitation spectrum
is similar to that at the ULS point, the entropies for block sizes 
that are multiples of three
can be well-fitted with the form given in Eq.~(\ref{eq:cardy}) 
with $c=2$, and for finite chains, a peak appears in $\tilde{s}(q)$ at
$q=2\pi/3$, in agreement with Refs.~\onlinecite{biqu} and \onlinecite{c_ls}.
Thus, 
peaks in the Fourier spectrum of the length-dependent block entropy
can provide useful information about QPTs, even when they scale to
zero in the thermodynamic limit.

We demonstrate this procedure 
on the example of the spin-one bilinear-biquadratic
model near the VBS point \cite{AKLT}, corresponding to
$\theta_{\text{VBS}} = \arctan 1/3\simeq 0.1024\pi$.
It is known \cite{schollwock} that this point is a 
disorder
point, where incommensurate
oscillations appear in the decaying correlation function;
however, the shift of the minimum of the  
static structure factor appears only at a larger  
$\theta_{\text{L}}=0.138\pi$, the so-called Lifshitz point. 
In earlier work [\onlinecite{legeza_qpt}], some of us showed that $s(N/2)$
has an extremum  as a function of 
$\theta$ at $\theta_{\text{VBS}}$. 
Here we show that this extremum is the indication that, in fact,
$\theta_{\text{VBS}}$ is a dividing point  
which separates regions with a different behavior of $s(l)$ and $\tilde{s}(q)$.

As usual in the DMRG approach, we consider open chains. The numerical 
calculations were performed using the dynamic block-state 
selection (DBSS) approach \cite{legeza_dbss}. 
The threshold value of the quantum 
information loss $\chi$ was set to $10^{-8}$ for the spin models
and to $10^{-4}$ for the fermionic model, and the upper cutoff on the
number of  
block state was set to $M_{\text{max}}=1500$. 

\begin{figure}[htb]
\includegraphics[scale=0.5]{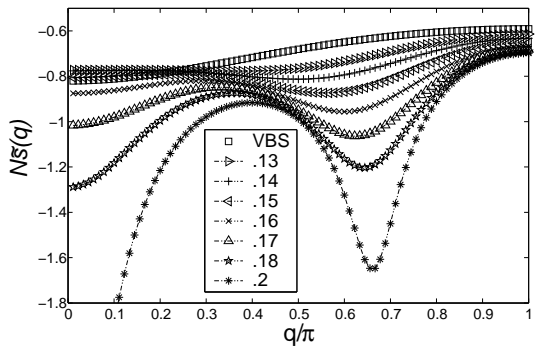}
\caption{Fourier-transformed entropy $\tilde{s}(q)$ for various $\theta$, obtained for a finite 
chain with $N=180$ lattice sites. The lines are guides to the eyes. 
The $q=0$ point, which has a large positive value, is not shown.}
\label{fig:biqu_incomm}
\end{figure}

At and below the VBS point, i.e., for $-\pi/4<\theta \leq
\theta_{\text{VBS}}$,  
$s(l)$ increases with $l$ for
small $l$, 
saturates due to the Haldane gap \cite{fan}, and then goes down to zero
again as $l$ approaches $N$. 
The Fourier spectrum $\tilde{s}(q)$ is a smooth function 
of $q$ (except for the $q=0$ component).
The transformed entropy $\tilde{s}(q)$ at the VBS point, depicted in
Fig.~\ref{fig:biqu_incomm}, illustrates this behavior.
For $\theta$ slightly larger than $\theta_{\text{VBS}}$, however, we
find that $s(l)$ does not increase to the saturation value purely
monotonically.
Instead, an incommensurate oscillation 
is superimposed.  
For somewhat larger $\theta$ values, $\theta>0.13\pi$, this oscillation 
persists in the saturated region, i.e., for blocks much longer
than the correlation length. 
This leads to a new peak 
in $\tilde{s}(q)$ which moves from small $q$ towards $q=2\pi/3$ as the ULS 
point is approached, and gets larger and narrower, as can be seen in
Fig.~\ref{fig:biqu_incomm}. 
This $\theta$ value is slightly smaller than, but close to, the Lifshitz point.

\begin{figure}[htb]
\includegraphics[scale=0.4]{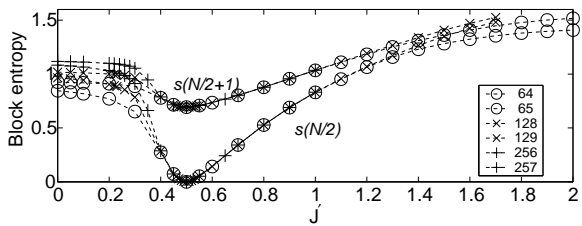} 
\caption{Entropy of blocks of length $N/2$ and $N/2+1$ as a function of 
$J'/J$ for the frustrated spin-1/2 Heisenberg chain for various chain
  lengths.
} 
\label{fig:majumdar_sz0}
\end{figure}

\begin{figure}[htb]
\includegraphics[scale=0.5]{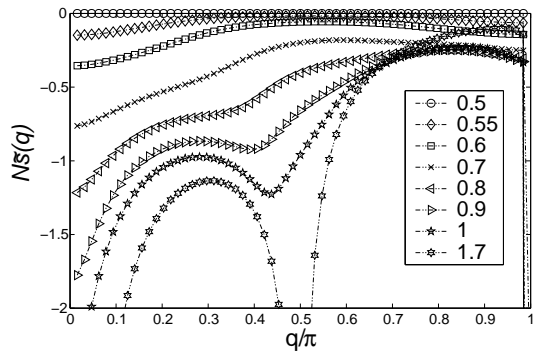} 
\caption{ Fourier-transformed entropy
$\tilde{s}(q)$ for various values of $J'J$ 
for the frustrated Heisenberg chain with $N=128$ lattice sites.} 
\label{fig:majumdar_fft_sz0}
\end{figure}

The spin-1/2 frustrated Heisenberg chain, Hamiltonian
(\ref{eq:ham_frust}), is also 
known to develop incommensurate correlations, for values of
$J^\prime/J > 0.5$ (the MG point).
As shown in Fig.~\ref{fig:majumdar_sz0}, the entropies of blocks of length
$N/2$ and $N/2+1$, although substantially different in value, both
display a minimum as a function of $J'/J$ at the MG point. 
Thus, the transition from commensurate (dimerized) to incommensurate 
correlations again is marked by an extremum of the block entropy. 
Similarly to the behavior in the 
neighborhood of the VBS point, in this model a long 
wavelength oscillation appears in $s(l)$ right above $J'/J > 0.5$,
which leads to a new peak in the 
Fourier spectrum.
The displacement with increasing $J^\prime/J$ is in this case,
however, towards $q=\pi/2$, as can be seen
in Fig.~\ref{fig:majumdar_fft_sz0}. This peak becomes larger and narrower for 
increasing $J'/J$, and again is a signature of incommensurability. 

\begin{figure}[htb]
\includegraphics[scale=0.5]{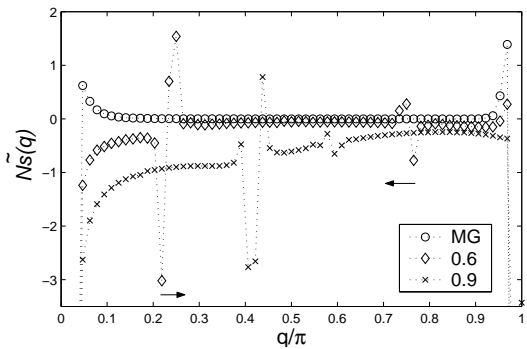}
\caption{Fourier-transformed entropy
$\tilde{s}(q)$ for various values of $J'/J$ for the 
triplet excited state
of the frustrated Heisenberg chain.
The arrows indicate the direction of movement of the peaks with
increasing $J^\prime/J$.
}
\label{fig:majumdar_fft_sz1}
\end{figure}

It is also interesting to examine the 
behavior of the block entropy in the lowest-lying triplet
excited states.
This is shown in Fourier-transformed representation, $\tilde{s}(q)$, in
Fig.~\ref{fig:majumdar_fft_sz1} for several $J'/J$ values.
At $J'=0.5J$, the peak at $q=\pi$ is displaced to
$q=\pi(1-1/N)$.
As $J'/J$ increases, we find two oppositely moving peaks.  
One appears exactly where the peak was found
for the ground state, while the other occurs at $\pi-q$. 
By also calculating the structure factor, S(q), we found that this second
peak is located at the same (J'/J-dependent) wave vector at which S(Q)
has its maximum \cite{white_affleck}. 
%
%
For Hamiltonian (\ref{eq:ham_biqu}), 
the block entropy of the first triplet state also has two oppositely
moving peaks.
One is again at the same location as the incommensurate peak in the
ground state, while the other is at $\pi-q/2$. 
Both peaks move towards $2\pi/3$ with increasing $\theta$. 

Having demonstrated the usefulness of studying the entropy profiles 
for models where the quantum critical points are known, we now use this 
procedure to study an commensurate-incommensurate
transition in the 1D $t-t'-U$ Hubbard model 
\begin{equation}   \begin{split}
      {\mathcal H} &= t \sum_{i\sigma} \left ( c^\dagger_{i\sigma} 
    c^{\phantom\dagger}_{i+1\sigma} + c^\dagger_{i+1\sigma} 
     c^{\phantom\dagger}_{i\sigma} \right )   \\ 
        &  \phantom{=,} 
      + t^\prime \sum_{i\sigma} \left ( c^\dagger_{i\sigma} 
    c^{\phantom\dagger}_{i+2\sigma} + c^\dagger_{i+2\sigma} 
     c^{\phantom\dagger}_{i\sigma} \right ) + U \sum_{i} n_{i\uparrow} n_{i\downarrow} \,, 
\label{eq:ham_nnh}
\end{split}   \end{equation}
which has been investigated recently \cite{japaridze}.
For the half-filled case (and setting $t=1$), the
competition between $t^\prime$ and the Coulomb
energy $U$ will determine whether the system is an insulator $(t^\prime < t^\prime_{\rm s})$
or a metal $(t^\prime > t^\prime_{\rm c})$.
For finite $U$ values, the transition
between these two states occurs in two steps. First the spin gap opens
at  $t^\prime_{\rm s}$ and then the charge gap closes at a larger
value $t^\prime_c$.
Between these two points, the wave vector becomes
incommensurate for $t^\prime>t^\prime_{\rm IC}$
-- the commensurate-incommensurate transition is independent of the 
metal-insulator transition \cite{japaridze}. 

\begin{figure}[htb]
\resizebox{7cm}{6.5cm}{
\includegraphics[scale=0.35]{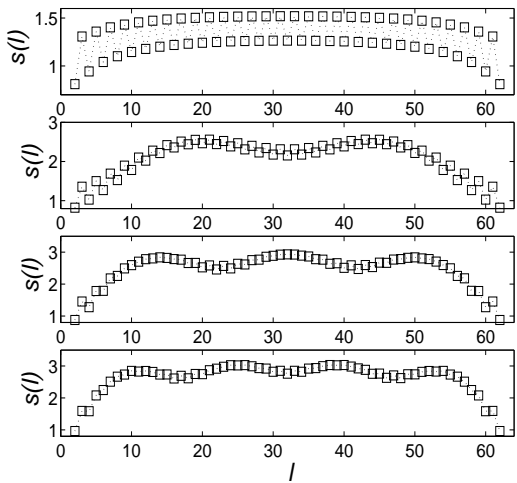}}
\caption{Block entropy profile of a finite chain with $N=64$ sites 
for the $t-t'-U$ Hubbard model for $U=3$ 
at
$t'=0.6, 0.625, 0.65, 0.675$ (from top to bottom).}
\label{fig:nnh_u3_block}
\end{figure}

As expected for a commensurate-incommensurate transition, we
find that the entropy of blocks of length $N/2$ and $N/2+1$  
display an extremum as a function of $t^\prime$. 
For very large $U$ values, where the model is equivalent to the
frustrated spin-1/2 Heisenberg chain, the extremum occurs at 
$t^\prime_{\rm IC}\simeq 1/\sqrt 2$, which maps to the MG point.
For $t^\prime>t^\prime_{\rm IC}$, an incommensurate oscillation in $s(l)$
becomes apparent. This behavior can be seen in Fig.~\ref{fig:nnh_u3_block} for
$U=3$, a value chosen
so that our results can be directly compared to those of 
Ref.~[\onlinecite{japaridze}]. 
When $\tilde{s}(q)$ is analyzed, it is found that a new peak appears
in the spectrum and again moves  
from small $q$ towards $q=\pi/2$ with the amplitude of $\tilde{s}(\pi)$ decreasing
with increasing $t^\prime$. 
Therefore, 
the commensurate-incommensurate phase boundary can be
easily determined by finding the extrema of $s(N/2)$ as a function of
$t^\prime$ for various $U$ values. 
The detailed entropy analysis of the full two-dimensional phase diagram 
will be
presented in subsequent work.

In conclusion, we have
shown that the 
length dependence of the 
block entropy and its Fourier spectrum, determined 
for finite systems, can be 
used to characterize phases in which the correlation function has an 
oscillatory character.
In addition, an extremum in the block entropy as a function of the
relevant model parameter,
which, in general, signals the appearance of or change in a
symmetry in the wave function, 
can also correspond to disordered points.
In this case, however, the entropy curve does not show 
anomalous behavior 
because this is not a phase transition in the conventional 
sense. 
When the decaying correlation function has an incommensurate oscillation, 
a new peak appeares close to $q=0$ in the Fourier spectrum and moves 
towards a commensurate 
wave vector as the control parameter is adjusted.
In the entropy of the triplet excited states, in addition to a
peak at the same position as in the ground-state entropy,
another peak appears at the wave vector
of the peak in the static structure factor.  
Our method is ideal for use in conjunction with the density-matrix 
renormalization-group algorithm because the block entropy
profile is generated as a by-product of the DMRG procedure.
This allows the
more difficult and sensitive calculation of 
correlation functions to be avoided. 

This research was supported in part by the Hungarian Research Fund (OTKA)
Grants No.\ T043330, F046356 and the J\'anos Bolyai Research Grant. 
The authors acknowledge
computational support from Dynaflex Ltd under Grant No. IgB-32.
\"O.~L.\ and L.~T.\ acknowledge useful discussions with G.\ F\'ath.


\begin{thebibliography}{99}


\bibitem{wu}
L.-A. Wu, M.S. Sarandy, and D.A. Lidar, Phys. Rev. Lett. {\bf 93}, 
250404 (2004). 

\bibitem{wootters}
W.K. Wootters, Phys. Rev. Lett. {\bf 80}, 2245 (1998). 

\bibitem{osborne}
T.J. Osborne and M.A. Nielsen, Phys. Rev. A {\bf 66}, 32110 (2002).

\bibitem{osterloh}
A. Osterloh, L. Amico, G. Falci, and R. Fazio, Nature {\bf 416}, 608 (2002).

\bibitem{syljuasen}
O.F. Sylju{\aa}sen, Phys. Rev. A {\bf 68}, 60301(R) (2003).

\bibitem{gu}
S.-J. Gu, H.-Q. Lin, and Y.-Q. Li, Phys. Rev. A {\bf 68}, 42330 (2003).

\bibitem{vidal}
J. Vidal, G. Palacios, and R. Mosseri, Phys. Rev. A {\bf 69}, 022107 (2004);
J. Vidal, R. Mosseri, and J. Dukelsky, {\sl ibid.} {\bf 69}, 054101 (2004).

\bibitem{roscilde}
T. Roscilde, P. Verrucchi, A. Fubini, S. Haas, and V. Tognetti, Phys. Rev.
     Lett. {\bf 93}, 167203 (2004); {\bf 94}, 147208 (2005).

\bibitem{yang}
M.-F. Yang, Phys. Rev. A. {\bf 71}, 30302(R) (2005).

\bibitem{zanardi}
P. Zanardi, Phys. Rev. A {\bf 65}, 42101 (2002).

\bibitem{gu2}
S.-J. Gu, S.-S. Deng, Y.-Q. Li, and H.-Q. Lin, Phys. Rev. Lett. {\bf 93},
86402 (2004).

\bibitem{larsson_single}
D. Larsson and H. Johannesson, Phys. Rev. Lett. {\bf 95}, 196406 (2005);
Phys. Rev. A {\bf 73}, 042320 (2006).


\bibitem{legeza_qpt}
{\"O}. Legeza and J. S\'olyom, Phys. Rev. Lett. {\bf 96}, 116401  (2006).

\bibitem{deng}
S.-S. Deng, S.-J. Gu, and H.-Q. Lin, Phys. Rev. B {\bf 74} 045103 (2006).

\bibitem{vidal_sl}
G. Vidal, J.I. Latorre, E. Rico, and A. Kitaev, Phys. Rev. Lett. {\bf 90}, 227902 (2003).
J.~I. Latorre, E. Rico, and G. Vidal, Quant. Inf. and Comp. {\bf 4},  48
  (2004).

\bibitem{korepin_sl}
V. E. Korepin, Phys. Rev. Lett. {\bf 92}, 096402  (2004).

\bibitem{holzhey} C. Holzhey, F. Larsen, and F. Wilczek, Nucl. Phys. {\bf B424}, 443 (1994).

\bibitem{cardy}
P. Calabrese and J. Cardy, J. Stat. Mech.: Theor. Exp. P06002 (2004).

\bibitem{affleck}
I. Affleck and A. W. W. Ludwig, Phys. Rev. Lett. {\bf 67}, 161 (1991).

\bibitem{white}
S.R. White, Phys. Rev. Lett. {\bf 69},  2863  (1992); Phys. Rev. B
         {\bf 48},  10345  (1993).

\bibitem{majumdar}
C. K. Majumdar and D. K. Ghosh, J. Mat. Phys. {\bf 10}, 1388 (1969); ibid 1399 (1969).

\bibitem{TB}
L. A. Takhtajan, Phys. Lett. A {\bf 87}, 479 (1982); H. M. Babujian, 
      Phys. Lett. A {\bf 90}, 479 (1982); 

\bibitem{ULS}
G. V. Uimin, JETP Lett. {\bf 12}, 225 (1970);
C. K. Lai, J. Math. Phys. {\bf 15}, 1675 (1974); 
B. Sutherland, Phys. Rev. B {\bf 12}, 3795 (1975).

\bibitem{biqu}
For its phase diagram see G. F\'ath and J. S\'olyom, Phys. Rev. B {\bf 44},
11836  (1991);  {\bf 47},  872  (1993); {\bf 51},  3620  (1995);


\bibitem{c_tb}
I. Affleck and F.D.M. Haldane, Phys. Rev. B {\bf 36}, 5291 (1987).

\bibitem{c_ls}
C. Itoi and M.-H. Kato, Phys. Rev. B {\bf 55}, 8295 (1997).

\bibitem{AKLT}
I.\ Affleck, T.\ Kennedy, E.\ Lieb, and H.\ Tasaki, Phys. Rev. Lett. {\bf 59},
799 (1987).

\bibitem{schollwock}
U. Schollwock, Th. Jolicoeur, and T. Garel, Phys. Rev. B {\bf 53}, 3304 (1996). 

\bibitem{legeza_dbss}
{\"O}. Legeza and J. S\'olyom, Phys. Rev. B {\bf 70},  205118  (2004).

\bibitem{fan}
H.\ Fan, V.\ Korepin, and V.\ Roychowdhury, Phys.\ Rev.\ Lett.~{\bf 93}, 227203 (2004).


\bibitem{white_affleck}
S.R.~White and I.~Affleck, \prb {\bf 54}, 9862 (1996).

\bibitem{japaridze}
G.I.~Japaridze, R.M.~Noack, and D.~Baeriswyl
(unpublished, cond-mat/0607054).


\end{thebibliography}
\end{document}